\documentclass[a4paper]{article}
\usepackage{bm}
\usepackage{authblk}
\usepackage{graphicx}
\usepackage{amsmath,amssymb}
\usepackage{amsfonts}
\usepackage{cite}
\usepackage{subfigure}
\title{Unbiased estimation of an optical loss at the ultimate quantum limit with twin-beams} 

\author[1,2]{Elena Losero}
\author[1,*]{Ivano Ruo-Berchera}
\author[1]{Alice Meda}
\author[1]{Alessio Avella}
\author[1,3]{Marco Genovese}
\affil[1]{INRIM, Strada delle Cacce 91, I-10135 Torino, Italy}
\affil[2]{DISAT, Politecnico di Torino, I-10129 Torino, Italy}
\affil[3]{INFN Sezione di Torino, via P. Giuria 1, I-10125, Torino, Italy}

\affil[*]{i.ruoberchera@inrim.it}

\begin{document}

\maketitle


\section*{Abstract}
Loss measurements are at the base of spectroscopy and imaging, thus permeating all the branches of science, from chemistry and biology to physics and material science.  However, quantum mechanics laws set the ultimate limit to the sensitivity, constrained by the probe mean energy. This can be the main source of uncertainty, for example when dealing with delicate system such as biological samples or photosensitive chemicals. It turns out  that ordinary (classical) probe beams, namely with Poissonian photon number distribution, are fundamentally inadequate to measure small losses with the highest sensitivity. Conversely, we demonstrate that a quantum-correlated pair of beams, known as “twin-beam state”, allows reaching the ultimate sensitivity for all energy regimes (even less than one photon per mode) with the simplest measurement strategy. One beam of the pair addresses the sample, while the second one is used as a reference to compensate both for classical drifts and for fluctuation at the most fundamental quantum level. This scheme is also absolute and accurate, since it self-compensates for unavoidable instability of the sources and detectors, which could otherwise lead to strongly biased results. Moreover, we report the best sensitivity per photon ever achieved in loss estimation experiments.

\section*{Introduction}

The measurement of changes in intensity or in phase of an electromagnetic field, after interacting with matter, is the most simple and effective way to extract relevant information on the properties of a system under investigation, whether a biological sample \cite{biology0, biology1} or a digital memory disc \cite{Gu2014}. Intensity measurements enable absorption/transmission estimation, the base of imaging and spectroscopy, pervasive and fundamental techniques in all science fields, from chemistry \cite{chemistry0} to  material science \cite{materialscience0} and physics\cite{astrophys0}. They are routinely employed in biomedical analysis \cite{medicine0,biomedical0, biomedical1}, as well as in atmospheric \cite{atmosphere0, atmosphere1, atmosphere2} and food sciences \cite{food0, food1}.

However, the optical transmission losses experienced by a probe beam while interacting with a system cannot be determined with arbitrary precision, even in principle. Quantum mechanics establishes fundamental bounds to the sensitivity \cite{Giovannetti2011,Demkowicz2015,pirandola2017,paris2007}, which is  limited, in general, by the mean energy of the probe, or, equivalently, by its mean number of photons .  This is in accordance to the intuitive idea that gaining the perfect knowledge on a system would require an infinite amount of physical resources.

The lower bound to the uncertainty, when restricted to the use of classical probe states, coincides with the one achieved by a coherent state, $U_{coh}\backsimeq[(1-\alpha)/\langle N_{P} \rangle]^{1/2}$\cite{pirandola2017}, where $\langle N_{P} \rangle$ is the mean number of photons of the probe and $0\leq\alpha\leq1$ is the loss of the sample. Indeed, this limit can be obtained in practice by any probe beam exibiting Poissonian photon statistics, as a laser beam (described theoretically by a coherent state) or even a thermal source like LEDs or incandescent light bulbs in the limit of extremely low photon number per mode. Note that the uncertainty depends on the loss parameter, and can be arbitrary small only in the asymptotic limit of high losses. For a faint loss, $\alpha\sim 0$, one retrieves the expression $U_{snl}= \langle N_{P} \rangle^{-1/2}$, usually referred as to "shot-noise-limit" (SNL).

Without restriction on the probe state, it has been shown \cite{paris2007,adesso2009} that the ultimate quantum limit (UQL) in the sensitivity for a single mode interrogation of the sample is  $U_{uql}\backsimeq\sqrt{\alpha}\, U_{coh}$, which scales much more favourably than the classical bound for small losses, a region which is particularly significant in many real applications.
It is worth noting that the use of quantum states does not improve the uncertainty scaling with the number of particles. This is different from what happens in phase shift estimation, in which a sensitivity scaling proportional to $\langle N_{P} \rangle^{-1}$ is reachable in ideal situations\cite{Giovannetti2011,Demkowicz2015}, the so called "Heisenberg limit". The fundamental difference is that phase shift is a unitary operation, preserving the purity of the state, while a loss is intrinsically non unitary. A loss can be  represented as the action of a beam splitter that mixes up the probe state in one port with the vacuum state in the other port, basically spoiling quantum features such as entanglement, which is necessary to approach the Heisenberg limit \cite{Demkowicz2015}.

It is known that single mode squeezed vacuum reaches $U_{uql}$  for small losses, $\alpha\sim 0$, and small number of photons $\langle N_{P} \rangle\sim 0$ \cite{paris2007}. Fock states $|N\rangle$, having by definition a fixed number of photon, approach $U_{uql}$  unconditionally, i.e. for all value of $\alpha$, but they cannot explore the regime of $\langle N_{P} \rangle < 1$ \cite{adesso2009}. The optimal performance of Fock states can be understood by considering that a loss can be easily estimated by comparing the number of photons of the probe before and after the interaction with the sample. The perfect knowledge of the photon number of the unperturbed Fock state allows one to detect better small deviations caused by the sample, which would remain hidden in the intrinsic photon number fluctuation of Poissonian distributed sources.

However it is challenging to produce experimentally true Fock states. A reasonable approximation of a Fock state with $N=1$ are the heralded single photons produced by spontaneous parametric down conversion (SPDC)\cite{Brida2011-0,Krapick2013}. In this process photons are always emitted in pairs with low probability, but one can get rid of the vacuum component since the detection of one photon of the pair heralds the presence of the other one. This scheme has been demonstrated recently for quantum enhanced absorption measurement both with post-selection of the heralded single photons \cite{Whittaker2017} and, more remarkably, with selection performed by active feed-forward enabled by an optical shutter \cite{sabines2017}.

Also quantum correlations of twin-beam (TWB) state  have shown the possibility of sub-SNL sensitivity in absorption/transmission measurement \cite{tapster1991,Hayat1999, Moreau2017,Brida2010,brida2011-1,Samantaray2017,MG15}, quantum enhanced sensing \cite{zhang2015,lopaeva2013, Pooser2015}, ghost imaging \cite{brida2011-2} and quantum reading of digital memories \cite{pirandola2011}. TWB states can be generated by SPDC \cite{review17} as well as by four wave mixing in atomic vapours \cite{Glorieux2011,Embrey2015,cao2017}, and expose a high level of quantum correlation in the photon number fluctuations between two corresponding modes, for example two propagation directions or two wavelengths. Even if super-Poissonian noise characterizes the photon distribution in one mode, the fluctuations are perfectly reproduced in time and space in the correlated mode. Sub-shot noise correlation of this state has been experimentally demonstrated both in the two-mode case \cite{Heidmann1987,Mertz1990,bondani2007,agafonov2011,iskhakov2016} and in the case of many spatial modes detected in parallel by the pixels of a CCD camera \cite{jedrkiewicz2004,brida2009,blanchet2008}. The exploitation of spatially multimode non-classical correlation has been proposed for high sensitivity imaging of distributed absorbing object \cite{gatti2008} and a proof of principle of the technique has been reported by Brida \textit{et al.} in 2010 \cite{Brida2010}. Recently our group has realized the first wide-field sub-SNL microscope \cite{Samantaray2017}, providing $10^{4}$ pixels images with a true (without post-selection) significant quantum enhancement, and a spatial resolution of few micrometers. This represents a considerable advancement towards a real application of quantum imaging and sensing.

The common idea behind these works is that the random intensity noise in the probe beam addressed to the sample can be known by measuring the correlated (reference) beam and subtracted.  Note that the two-beams approach is extensively used in standard devices like spectrophotometers, where a classical beam is split in two by a beam splitter and one beam is used to monitor the instability of the source and detectors and to compensate for them.  This is particularly effective in practical applications, since unavoidable drifts in the source emission or detector response would lead to strong bias, especially in the estimation of small absorptions. However, in classical correlated beams (CCB) generated in this way, only the super-Poissonian component of the fluctuations is correlated (sometimes called classical "excess noise"),  whereas the shot noise remains uncorrelated and cannot be compensated. Therefore TWB represent the natural extension to the two-beam approach to the quantum domain, promising to be especially effective for small absorption measurement and when low photon flux is required.


It has been theoretically demonstrated \cite{illuminati} that  using TWB for loss estimation the UQL is in principle attainable; nevertheless the existence of an experimental estimator fit for this purpose is still an open question, as it is its explicit expression .

Here, we show that the answer to this question is unconditionally positive, for all the energy regime and all values of the loss parameter $\alpha$. Therefore, TWB overcome the limitations of both single mode squeezed vacuum and Fock states, representing the practical best choice for pure loss estimation.
We prove this result by an operative approach: we consider a specific and simple measurement strategy that is to evaluate the ratio between the photon number measured in the probe and in the reference beam. In the ideal lossless detection case this is sufficient to reach the ultimate quantum limit.
Taking into account for experimental imperfections, we derive the uncertainty advantage of the twin-beam with respect to the single classical beam (SCB)  and to the CCB  case. This quantum enhancement  can be expressed in terms of experimental parameters related to the "local" photon statistics of the two beams separately, and the amount of non-classical correlation of the joint photon number statistics.

In a recent work \cite{Moreau2017}, a different optimized estimator which allows improving the sensitivity in case of strong non-ideal quantum efficiencies has been proposed. The drawback is that this method requires the accurate and absolute characterization of the measurement apparatus, in particular the absolute values of the quantum efficiencies of the detectors and of the excess noise of the source. This aspect places a strong practical limitation, because the determination of quantum efficiency, especially at the few photon level, with uncertainty less than $10^{-3}$ is extremely challenging, limiting the overall accuracy of the method; then, instabilities could also affect the measurement. We show that our estimator behaves almost as good as the optimized one for relatively high values of the efficiencies (the condition of our experiment), but it requires the weakest assumptions on the stationarity of the system and does not require absolute value of any parameter.

Finally we perform the experiment, measuring intensity correlations in the far field of multi-mode parametric down conversion by a standard low noise and high efficiency (95\%) CCD camera. For a sample loss of $\sim 2\%$, we report an experimental quantum enhancement in the estimation  uncertainty of $1.51 \pm 0.13 $ with respect to the single beam classical probe and of $2.00 \pm 0.16$  compared to the classical two-beam approach. The enhancement is preserved up to a sample loss of $\sim70\%$. It represents, in our knowledge, the best sensitivity per photon ever reported up to now in a loss measurement.

\section*{Theory}

\begin{figure}[htbp]
	\centering
	\includegraphics[trim= 0 2cm 0 2cm, clip=true, width=0.7\textwidth]{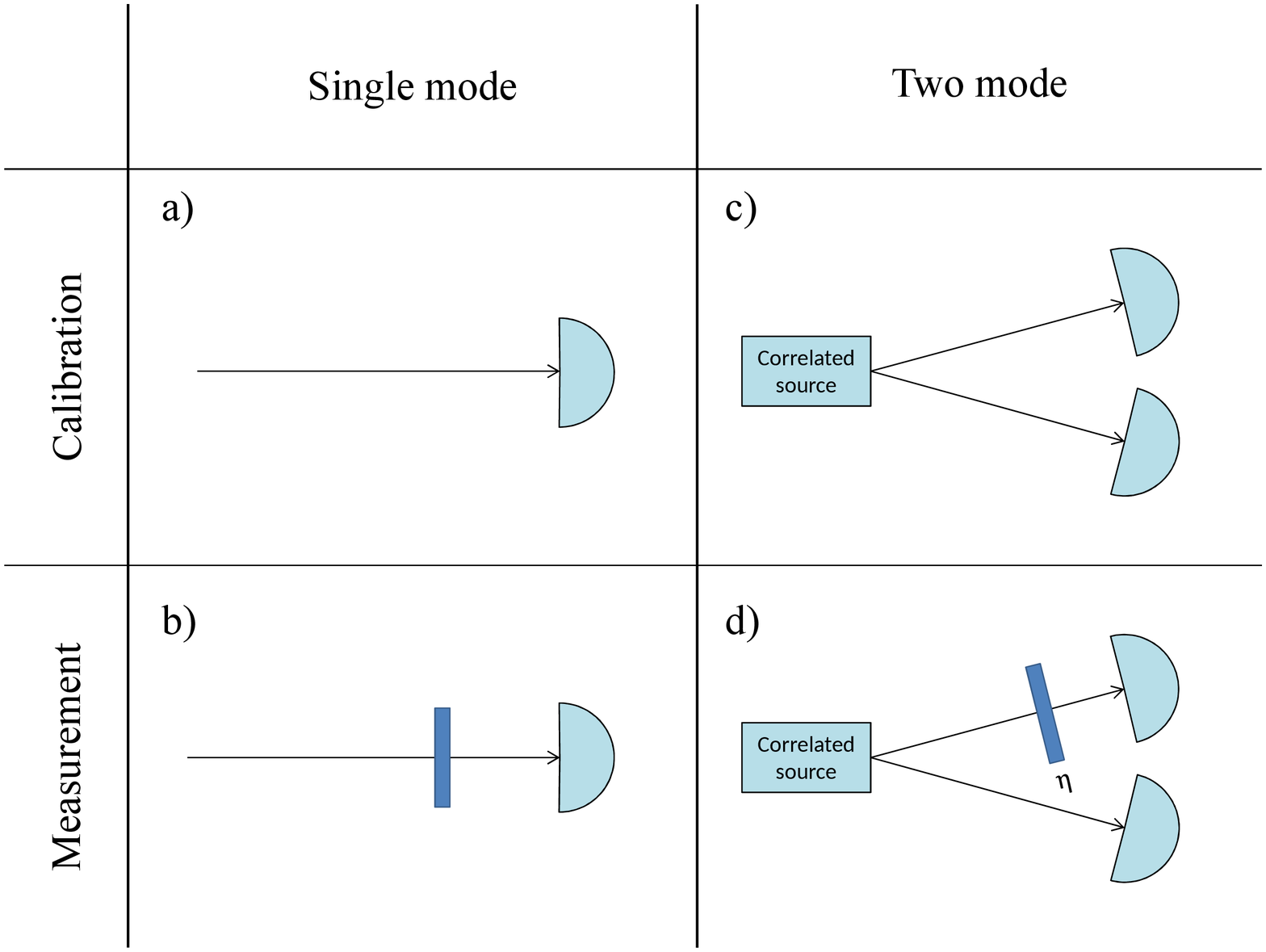}
	\caption{Two possible schemes to estimate the absorption coefficient $\alpha$. In the single-mode case (a) and (b) there is no correlation between probe and reference beam, i.e. $\langle \Delta N_p \Delta N_R \rangle =0$ while in the two-mode case (c) and (d) $\langle \Delta N_p \Delta N_R \rangle \neq 0$. Different possibilities of input states for both the schemes are discussed in the text.}\label{dr-diff}
\end{figure}

In practice, an optical loss $\alpha$ can be easily measured by comparing the number of photons of the probe $N'_{P}$  after a lossy interaction, with a reference value $N_{R}$, which can be evaluated in a previous moment in absence of the sample (Fig. \ref{dr-diff}a) or by the help of a second beam (Fig. \ref{dr-diff}d). In particular, one can define the estimator
\begin{equation} \label{S_alpha}
S_{\alpha}= 1-\gamma \frac{N'_{P}}{N_{R}} .
\end{equation}
The factor $\gamma=\langle N_{R}\rangle/ \langle N_{P}\rangle$ should be introduced in case of unbalancing between the mean energy of probe and reference beams and evaluated in a  pre-calibration phase of the apparatus (Fig. \ref{dr-diff}c). A loss is a random process modelled by the action of a beam splitter of transmission $1-\alpha$, so that the statistics of the photon counting of the probe beam is modified in this way \cite{review17}:
\begin{align}
& \langle N'_{P} \rangle = (1-\alpha) \langle N_{p} \rangle, \label{mean_Np} \\
& \langle \Delta^2 N'_{P} \rangle = [(1-\alpha)^2(F_{p}-1)+1-\alpha]. \label{var_Np}
\end{align}
Here $N_{p}$ is the  measured photon number without the sample. Its fluctuation is represented by the Fano factor $F_{p} = \langle \Delta^2 N_{p}\rangle /\langle N_{p} \rangle \geq 0$ which quantifies the non-classicality of the photon statistics. In particular $F<1$ indicates sub-Poissonian noise \cite{fano} and in general the possibility to surpass the SNL.

The mean value of the absorption is evaluated according to the definition as  $\langle S_{\alpha}\rangle=1-\gamma \langle N'_{P}\rangle/\langle N_{R}\rangle$ and inserting the mean value reported in Eq. (\ref{mean_Np}) leads to the unbiased estimation $\langle S_{\alpha}\rangle=\alpha$.

By propagating the uncertainty of the quantities $N'_{P}$ and $N_{R}$ on $S_{\alpha}$, and rewriting the variance $\langle \Delta^{2} N'_{p} \rangle$ in terms of the unperturbed one $\langle \Delta^{2} N_{p} \rangle$ the quantum expectation value of fluctuation is:
\begin{equation} \label{Ualpha}
\Delta^2 S_{\alpha}\backsimeq U_{uql}^{2}+ \frac{(1-\alpha)^{2}}{ \langle N_P \rangle} \frac{2\sigma_{\gamma}}{\gamma}.
\end{equation}
The most relevant quantity appearing in Eq.(\ref{Ualpha}) is the positive factor:
\begin{equation} \label{sigma_gamma}
\sigma_{\gamma}= \frac{\langle \Delta^2(N_{R}- \gamma N_{P})\rangle }{ \langle N_{R}+ \gamma N_{P} \rangle}= \frac{\langle \Delta^2 N_{R}\rangle+\gamma^{2}\langle \Delta^2 N_{P}\rangle-2\gamma\langle \Delta N_{P} \Delta N_{R}\rangle}{ \langle N_{R}+ \gamma N_{P} \rangle}.
\end{equation}
In the case of $\gamma=1$ it represents the quantifier of the non-classical correlation known as noise reduction factor (NRF), $\sigma=\sigma_{\gamma=1}$, where the bound between classical and quantum correlations is set by $\sigma=1$. Thus, the uncertainty is expressed in terms of simple measurable quantities related to the photon number statistics, i.e. the intensity fluctuations. Eq. (\ref{Ualpha}) shows that whenever $\gamma=1$ and  $\sigma=0$  the  UQL is retrieved, $\Delta^2 S_{\alpha}(\gamma=1,\sigma=0)\backsimeq U_{uql}^{2}$.

In the following we consider different states for the probe and the reference beam to establish the limit to the sensitivity in relevant scenarios.

Let us first focus on the states which do not present correlation between probe and reference (e.g. the measurements on the probe and reference beam are performed in two different moments, refer to Fig.\ref{dr-diff}a and b), so that $\langle \Delta N_{P} \Delta N_{R}\rangle=0$ .

\begin{itemize}
	\item \textit{Fock states}.  It is clear that the only chance for uncorrelated states to achieve the condition $\sigma_{\gamma}=0$ and hence the UQL according to Eq. (\ref{Ualpha}) is to have null fluctuation in the photon number both for the reference and probe beam, $\langle\Delta^2 N_{R}\rangle\equiv\langle\Delta^2 N_{P}\rangle\equiv0$. This means that the state must be the product of two Fock states, $ |N\rangle_{P}\bigotimes|N\rangle_{R}$. Thus, as anticipated, unperturbed Fock states reaches the UQL unconditionally, i.e.  for all the value of the parameter, with the only limitation that the mean photon number cannot be arbitrarily small \cite{adesso2009} (i.e.$\langle N_P \rangle \geq 1$).
		\begin{equation} \label{Ualpha_fock}
	\Delta^2 S_{\alpha}^{(Fock)}\backsimeq  U_{uql}^{2}
	\end{equation}
	
	\item \textit{Coherent states}.  Let us now consider the state  $ |coh\rangle_{P}\bigotimes|coh\rangle_{R}$, particularly interesting for its simple experimental implementation. In the photon number basis, coherent states have the form  $|coh\rangle=e^{-\frac{1}{2}\langle n \rangle}\sum_{n=0}^{\infty}  \frac{\langle n\rangle^{n/2}}{\sqrt{n!} }|n\rangle$, following the Poissonian photon number distribution $P_{coh}(n)=e^{-\langle n \rangle} \langle n \rangle^{n}/n!$, which has the property $\langle\Delta^2 n\rangle=\langle n\rangle$. Thus, substituting the variances with the mean values in the right hand side of Eq. (\ref{sigma_gamma}) one get  $\sigma_{\gamma}=(1+\gamma)/2$, and accordingly:
	\begin{equation} \label{Ualpha_coh}
	\Delta^2 S_{\alpha}^{(coh)}\backsimeq  U_{uql}^{2}+ \frac{(1-\alpha)^{2}}{ \langle N_P \rangle} \frac{1+\gamma}{\gamma}.
	\end{equation}
    The lower limit for a pair of coherent states is reached under the condition of $\gamma\gg1$, i.e. when the reference beam has much more energy than the transmitted probe, and the relative fluctuation on its photon number becomes negligible. In this case one gets $\Delta^2 S_{\alpha}^{(coh)}= (1-\alpha)/ \langle N_P \rangle = U_{coh}^2$. In practice, one can also consider an equivalent situation, in which the reference uncertainty has been statistically reduced to a negligible contribution by a long acquisition time in the calibration phase  (Fig. \ref{dr-diff}a), namely a time much longer than the one used for the measurement of the probe beam in presence of the sample (Fig. \ref{dr-diff}b). Indeed, replacing the variable $N_R$ with its mean value  $\langle N_R\rangle$ in the definition of $S_{\alpha}$ and of $\sigma_{\gamma}$ in Eq. (\ref{sigma_gamma}) leads to the an identical  limit of the sensitivity. 	
\end{itemize}

More in general, it is convenient to rewrite the noise reduction factor for uncorrelated states in terms of the measurable Fano factor of each beam in absence of the sample, i.e.  $\sigma_{\gamma}= (F_{R}+\gamma F_{P})/2$. With this substitution,  Eq.(\ref{Ualpha}) becomes:
\begin{equation} \label{Ualpha_Uncorr}
\Delta^2 S^{(unc)}_{\alpha}\backsimeq U^{2}_{uql}+\frac{(1-\alpha)^{2} }{ \langle N_P \rangle} \left(\frac{1}{\gamma} F_{R}+ F_{P}\right) .
\end{equation}

The measured Fano factors account for the statistics of light sources and for detection losses and inefficiency. If $0\leq\eta_{j}\leq1 (j=P,R)$ is the overall detection efficiency for each single beam, the measured Fano factor can be written as $F_{j}= \eta_{j} F^{(0)}_{j}+1-\eta_{j}$, where $F^{(0)}_{j}$ refers to the one of the unperturbed state of the source. As expected, detection losses deteriorate the non classical signature of the probe and reference beams, preventing the real possibility to reach the UQL even with Fock states.

Considering now joint states where a correlation between probe and reference is present, i.e. $\langle \Delta N_{P} \Delta N_{R}\rangle\neq0$ (Fig. \ref{dr-diff}c and d) we have:

\begin{itemize}
	\item \textit{TWB state}. Two mode twin beam state, generated by SPDC, is represented by the following entangled state in the photon number basis $\{ |n\rangle \}$ \cite{squeezed}:
\begin{equation} \label{TWB}	
	|TWB\rangle_{PR}=[\langle n \rangle+1]^{-1/2}\sum_{n=0}^{\infty}\left[\frac{\langle n \rangle }{\langle n \rangle+1}\right]^{n/2}|n\rangle_P|n\rangle_R.
\end{equation}
	
The two modes, separately, obey to a thermal statistics each, where $\langle\Delta^2 n\rangle=\langle n\rangle(1+\langle n\rangle)$. However, they are balanced in the mean energy, $ \langle n_P \rangle=\langle n_R \rangle$, and their fluctuations are perfectly correlated,  $\langle\Delta n_P \Delta n_R\rangle=\langle\Delta^2 n\rangle$. This leads to $\gamma=1$ and $\sigma=0$, thus demonstrating that unperturbed TWB reach the $U_{uql}$, according to Eq. (\ref{Ualpha}). Note that this result is independent on the value of the parameter $\alpha$ and on the energy of the probe beam which can contain less than one photon per mode on average. Indeed, this is usually the case in experiments.

	\item \textit{Classical correlated beams (CCB)}. Let us consider a bipartite correlated state produced by a unitary splitting of a single beam. Given a splitting ratio $0\leq\tau\leq1$, it turns out that the statistics of the two out-coming beams, the probe and the reference, is characterized by $\gamma=\tau^{-1}-1$  and $\sigma_{\gamma}=(2\tau)^{-1}$, which are remarkably independent on the photon number distribution of the initial beam. Substituting these values in Eq. (\ref{Ualpha}) leads to the same uncertainty of two uncorrelated coherent beams $\Delta^2 S^{(CCB)}_{\alpha}=\Delta^2 S^{(coh)}_{\alpha}$, reported in Eq. (\ref{Ualpha_coh}). It shows that classical correlation can never approach the UQL, and that the lower uncertainty is achieved for a splitting ratio $\tau\backsimeq 0$ corresponding to a strong unbalancing of beam energies, $ \langle n_P \rangle\ll\langle n_R \rangle$. Therefore, for the specific measurement strategy considered here and whatever the input state, it is convenient to use a highly populated reference beam and a weak prope beam. This result is in agreement with the behaviour reported by Spedalieri \textit{et al.} \cite{spedalieri} in the complementary situation in which the input state is a thermal one while the measurement strategy is the most general one  allowed by quantum mechanics.
	
\end{itemize}

Finally, to better understand how losses or excess noise of the source influence the final accuracy in real experiment we note that the parameter $\sigma_{\gamma}$ can be rewritten as $\sigma_{\gamma}= \frac{\left(\gamma+1\right)}{2} \sigma + \frac{\left(\gamma-1\right)}{2}(F_{R}-\gamma F_{P})$. In presence of equal losses in both the branches $\eta_{R}=\eta_{P}=\eta$, the noise reduction factor, expressed in terms of the ideal unperturbed one $\sigma^{(0)}$, is  $\sigma=\eta \sigma^{(0)}+1-\eta$. For the relevant case of a TWB state, assuming the same detection efficiency in the two channels, it is $F_{R}=F_{P}$, $\gamma=1$ and $\sigma^{(0)}=0$, leading to:
\begin{equation} \label{Ualpha_TWB}
\Delta^2 S^{(TWB)}_{\alpha, \eta}\backsimeq U^{2}_{uql}+2\frac{(1-\alpha)^{2} }{ \langle N_P \rangle} \left(1-\eta\right).
\end{equation}

This expression shows how the degradation of the accuracy in presence of losses prevents reaching the UQL in practice.

On the other side, for $\gamma=1$, balanced CCB (bCCB) fulfills the lower classical bound $\sigma_{\gamma}= \sigma= \sigma^{(0)}=1$, thus using Eq. (\ref{Ualpha}) we obtain:	 	
\begin{equation} \label{Ualpha_bCCB}
\Delta^2 S^{(bCCB)}_{\alpha, \eta}\backsimeq U^{2}_{uql}+2\frac{(1-\alpha)^{2} }{ \langle N_P \rangle}=\frac{(1-\alpha)(2-\alpha)}{\langle N_P \rangle}.
\end{equation}
	 	
Note that in case of bCCB, the accuracy is immune from the detection losses but it is always worse than in the case of TWB reported in Eq.(\ref{Ualpha_TWB}).	

Up to now we have analyzed the performance of the specific estimator in Eq. (\ref{S_alpha}), showing that it allows reaching the optimal limits both for classical and quantum states, in particular using TWB state the UQL is retrieved. However, other estimators have been considered in literature for absorption measurement with TWB.
An interesting alternative is the estimator used in the recent experiment by Moreau \textit{et al.}\cite{Moreau2017},
\begin{equation}\label{opt}
S'_\alpha = 1 - \frac{N'_P - k \Delta N_R+ \delta E} {\langle N_P \rangle},
\end{equation}

where the weight factor $k$ can be determined in order to minimize the uncertainty on $S'_\alpha$, while $\delta E$ is a small correction introduced to render the estimator unbiased.  However, $k$ and $\delta E$ need to be estimated in a phase of pre-calibration of the apparatus. In particular it turns out that  $k_{opt}$ is a function of the detection efficiencies of the channels and the local excess noise $k_{opt}=f(\eta_P, \eta_R,F_P,F_R)$ while $\delta E$ depends also from the measured covariance $\langle \Delta N_{P} \Delta N_{R}\rangle$. We have evaluated analytically in the general case, with the only hypothesis of balanced sources, the expected uncertainty of the estimator in Eq. (\ref{opt}) when $k=k_{opt}$. For the sake of simplicity, here we report the expression obtained in case of symmetric statistical properties of the channels, $\gamma = 1$ and $F_P=F_R=F$:
\begin{equation} \label{Ualpha_opt}
\Delta^2 S'_{ \alpha} =U^{2}_{uql}+\frac{(1-\alpha)^{2} }{ \langle N_P \rangle} \sigma\left(2-\frac{\sigma}{F}\right).
\end{equation}

For TWB and no-detection losses, the noise reduction factor $\sigma$ is identically null and the UQL is retrieved also with this estimator. Taking into account balanced detection losses, and the common experimental case of a mean photon number per mode much smaller than one, one can substitute in Eq.(\ref{Ualpha_opt})  $\sigma=1-\eta$ and $F\backsimeq1$. Therefore, the uncertainty becomes:
\begin{equation} \label{Ualpha_opt_TWB}
\Delta^2 S'^{(TWB)}_{ \alpha, \eta} =U^{2}_{uql}+\frac{(1-\alpha)^{2} }{ \langle N_P \rangle}  \left(1-\eta^2\right).
\end{equation}
Comparing the uncertainty in Eq. (\ref{Ualpha_opt_TWB}) with the one reported in Eq. (\ref{Ualpha_TWB}) makes it clear that the estimator $S'_\alpha$ proposed in \cite{Moreau2017} performs better than $S_\alpha$, especially when detection losses are considerable. 

Finally, in Brambilla \textit{et al.}\cite{gatti2008}  it is suggested to measure the absorption by a differential measurement, considering the following estimator:
\begin{equation}\label{S2}
S''_\alpha= \frac{N_R-\gamma N'_P}{\langle N_R\rangle}.
\end{equation}
Assuming a source producing a pairs of beams with the same local statistical properties, the  variance of $S''_\alpha$ can be calculated as:
\begin{equation}\label{aaa}
\Delta^{2} S''_{\alpha} = \frac{[ 2(1-\alpha)\sigma_{\gamma}+\alpha + (F_R-1)\alpha^ 2]}{\gamma \langle N_P \rangle}.
\end{equation}
However, this choice is not optimal and depends on the value of the measured local statistics: in the best case of unperturbed TWB, in which $\sigma_{\gamma}=0$ and $\gamma=1$, it approaches $U_{uql}$ only asymptotically for $F_{R} \alpha^{2}\sim 0$. In TWB, produced experimentally by SPDC, the statistics of each mode is thermal with a photon number per mode much smaller than one,  thus $F_{R}\backsimeq1$ and the condition reduces to $\alpha\sim 0 $. Conversely, for high value of the estimated losses, $\alpha\sim 1 $, the performance of this estimator is much worse than the one of $S_{\alpha}$ and $S'_{\alpha}$.

\section*{Experiment}

\begin{figure}[htbp]
	\centering
	\includegraphics[ trim= 3.5cm 2cm 2 0cm, clip=true, width=0.4\textwidth, angle=270 ]{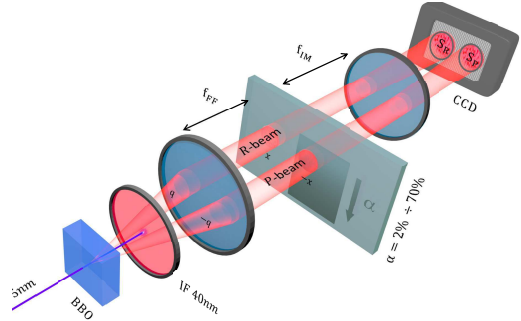}
	\caption{Scheme of our experimental set-up. In the BBO crystal two beams with perfect correlation in the photon number (TWB state) are generated. The probe beam passes trought the sample and is then detected in the $S_P$ region of the CCD, on the contrary the reference beam goes directly to $S_R$, without interacting with the sample. A detailed description of the optical components can be found in the text.}\label{setup}
\end{figure}

A scheme of the experimental set-up is reported in Fig. \ref{setup}.

A CW laser-beam (10mW at $\lambda_{pump}=405 nm)$ pumps a 1cm Type-II-Beta-Barium-Borate (BBO) non linear crystal, where SPDC occurs and two beams with perfect correlation in the photon number are generated. Note that the state $|\Psi \rangle$ produced by SPDC process is intrinsically multi-mode 
and can be expressed, in the plane-wave pump approximation, as a tensor product of two-modes TWB states of the form in Eq. (\ref{TWB}) as: $|\Psi \rangle = \otimes_{\bf{q},\lambda} | TWB \rangle _{\bf{q}, \lambda}$, where $\bf{q}$ and $\lambda$ are respectively the transverse momentum and the wavelength of one of the two photons produced, while momentum and wavelength of the other photon are fixed by energy and momentum conservation.

The far field of the emission is realized at the focal plane of a lens with $f_{FF}=1 cm$ focal length. Then a second lens, with $f_{IM}=1.6 cm$, images the far field plane to the detection plane. The magnification factor is M=7.8.
The detector is a charge-coupled-device (CCD) camera Princeton Inst. Pixis 400BR Excelon operating in linear mode. It presents high quantum efficiency ($>95\%$ at 810nm), 100$\%$ fill factor and low noise (read-noise has been estimated around 5 $e^-/(pixel \cdot second)$). The physical pixel of the camera measures 13 $\mu m$, nevertheless, not being interested in resolution, we group them by 24x24 hardware binning. This allows us to reduce the acquisition time and the effects of the read-out noise.
 Just after the crystal an interference filter ($800 \pm 20 nm, 99\%$ transmittance) is positioned to select only the modes of frequencies around the degeneracy, $\lambda_d=2 \lambda_{pump}$. This choice allows the presence of different spatial modes, in our case we have $M_{sp} \sim 2500$ spatial modes impinging on each detection area, $S_P$ and $S_R$, where $P$ and $R$ subscripts refer to the probe and reference beam, respectively. We integrate the signals in $S_R$ and in $S_P$. The sample consists in a coated glass-slide with a deposition of variable absorption coefficient $\alpha$ intercepting the probe beam the focal plane. We consider values of $\alpha$ from $1\%$ to $70\%$.  Finally, in order to check the theoretical model at varying $\eta_R$ and $\eta_P$, neutral filters of different absorption can be eventually positioned on the beams path.

The acquisition time of a single frame is set to $100ms$, whilst the coherence time of the SPDC process is around $10^{-12}s$, thus the number of the detected temporal modes is approximatively $M_{t} \sim  \cdot 10^11$. Since in each detection area we register around $50 \cdot 10^4$ photons per frame, it follows that the occupation number of the single spatio-temporal mode is $\mu \sim 2 \cdot 10^{-9}$ photons/mode. Being $\mu \ll 1$, this implies that the statistic of a single mode is well modelled by a Poissonian statistic: it follows that if only one beam is considered the measurements are shot-noise limited.

However, it is possible to go beyond the shot noise limit exploiting the photon number correlation between pairs of correlated modes.
In the plane wave pump approximation with transverse momentum $\bm{q}_{pump}=0$, in the far field region any mode with transverse momentum $\textbf{q}$ is associated with a single position $\textbf{x}$ according to the relation: $\textbf{x}={\frac{2 c f_{FF} }{ \omega_{pump}}} \textbf{q}$, where $c$ is the speed of light, $f_{FF}$ the focal length of the first lens and $\omega_{pump}$ the laser frequency.
The exact phase-matching condition for correlated modes
$\bm{q}_P + \bm{q}_R= \bm{q}_{pump}=0$
becomes in the far field, for degenerate wavelengths  $\lambda_P=\lambda_R=2\lambda_{pump}$,  a condition on their position: $\bm{x}_P + \bm{x}_R=0$.
Under the hypothesis of plane wave pump it is therefore expected that two symmetric pixels of the camera, respect  to  the  pump direction, always detect the same number of photons.
For a realistic pump with a certain spread $\Delta \bm{q}$ it follows: $\bm{x}_P + \bm{x}_R= 0 \pm \Delta \bm{x}= \pm {\frac{2 c f_{FF}}{\omega_{pump}}}\Delta \bm{q}$. 
$\Delta \bm{x}$ represents the size in the far field of the so called coherence area, $A_{coh}$, area in which photons from correlated modes can be collected. Moreover, the non-null frequency bandwidth (about 40 nm in our experiment) determines a further broadening of the spot in
which correlated detection events occur. To experimentally  measure the size of $A_{coh}$ the spatial cross-correlation between the two beams can be considered \cite{Samantaray2017}. 
Its evaluation is important to compare it with the detection area $A_{det}$ since, to detect a significant level of correlation, it is necessary that $A_{det} \geq A_{coh} $. In our case, integrating on the two regions of interest this condition is fully fulfilled, indeed it holds $A_{det} \gg A_{coh}$.
In general the measured NRF can be modelled as \cite{model}: $\sigma_{\gamma}= \frac{1+\gamma}{2}-\eta_R \eta_{coll} \geq 0$, where two contributions are present.
\begin{itemize}
	\item $0 \leq \eta_R \leq 1$, the detection efficiencies of the reference optical path
	\item $0 \leq \eta_{coll} \leq 1$, the collection efficiency of correlated photons. This factor depends on our ability in defining correlated regions  and reasonably increases enlarging the dimension of the detection area.
\end{itemize}
In our experimental situation, since $A_{det} \gg A_{coh}$ it follows $\eta_{coll} \rightarrow 1$ and consequently $\sigma_{\gamma} = \frac{1+\gamma}{2}- \eta_R$. Inverting this relation offers a useful way to measure the  absolute efficiencies on the two channels, as discussed in \cite{brida2010OE}. In the experimental situation corresponding to Fig.\ref{graph} we measured $ \sigma_{\gamma}=(0.24 \pm 0.03)$ and $\gamma = 1.006$, which implies detection efficiencies $\eta_R = \eta_P = 0.76 $, as reported in the caption. The same method has been adopted to evaluate the efficiencies in the other cases, reported in Fig. \ref{prova3lim} and \ref{prova4lim}.

In all these figures the mean values of $\alpha$ (x-axis) and their corresponding uncertainties $\Delta \alpha$ (y-axis) have been obtained acquiring 200 frames with the absorbing sample inserted. Repeating each measurement 10 times the error bars have been estimated.
In particular, for each frame, we integrate the data on $S_R$ and $S_P$, opportunely corrected for the background, obtaining $N_R$ and $N'_P$, necessary for the estimation of the mean absorption $\alpha$  according to the different estimators considered, in  Eqs.(\ref{S_alpha}-\ref{opt}-\ref{S2}).

To reproduce the single-mode classical strategy we performed a calibration measurement without the sample obtaing $\langle N_R \rangle$; we then estimate $\alpha$ as:
\begin{equation}
S_{\alpha}^{(unc)}=1-\gamma \frac{N'_P}{\langle N_R \rangle}
\end{equation}
As discussed in the theory section, this strategy leads, in the ideal situation, to the classical lower limit $U_{coh}$.

Finally to reproduce the bCCB case we consider a different region of the detector $S'_R$, displaced from $S_R$ and only classically correlated with $S_P$.

Note that from the calibration measurement also $\gamma, \sigma_{\gamma}, F_P$ and $F_R$ can be simply evalueted.


\section*{Results and Discussion}

\begin{figure}[htbp]
	\centering
	\includegraphics[trim= 0 4cm 0 4cm, clip=true, width=0.8\textwidth]{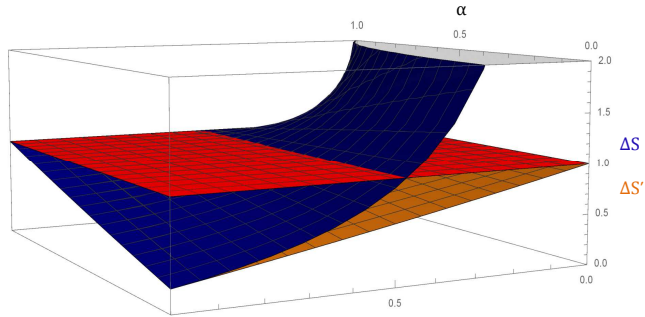}
	\caption{Uncertainty on $\alpha$, normalized to the single mode classical case ($\Delta S_{\alpha}^{(unc)}$, red surface),  using TWB as input state and the two different estimators presented in the text ($S_{\alpha}$ in Eq.(\ref{S_alpha}), blue surface,  and $S$'$_{\alpha}$ in Eq.(\ref{opt}), orange surface) in function of the losses on the reference path, $\eta_R$, and $\alpha$. It turns out that for $\eta_R$ sufficiently close to one  $\Delta S_{\alpha}^{(TWB)} \sim \Delta S$'$_{\alpha}^{(TWB)}$; on the other hand for, $\eta_R < 0.5$, $\Delta S_{\alpha}^{(TWB)} > \Delta S_{\alpha}^{(unc)}$ while $\Delta S$'$_{\alpha}^{(TWB)}$ always remains below this limit.}\label{gain3d}
\end{figure}

In Eq. (\ref{Ualpha_TWB}) and (\ref{Ualpha_opt_TWB}) we have explicitly reported the uncertainty achieved by TWB for two different estimators in case of matched detection efficiencies in the probe and reference beam. The unbalanced case leads to cumbersome analytical expressions, so we report this situation graphically  in Fig. \ref{gain3d}.  The uncertainties on $S_{\alpha}^{(TWB)}$ and $S$'$_{\alpha}^{(TWB)}$ are compared when varying $\eta_R$ at fixed $\eta_P$.
It emerges that for $\eta_R=1$ the two estimators offer exactly the same quantum enhancement, maximum for $\alpha \ll 1$. Nonetheless, for $\eta_R \neq 1$ and sufficiently large, the performances of the two estimators remain comparable. Instead when $\eta_R < 0.5$ the uncertainty on $S_{\alpha}^{(TWB)}$ becomes greater than the one classically achievable considering one single beam $\Delta S_{\alpha}^{(unc)}$; on the contrary $\Delta S$'$_{\alpha}^{(TWB)}$ maintains always below $\Delta S_{\alpha}^{(unc)}$.
Note that In Fig.\ref{gain3d} we fix $\eta_P=0.76$ (the value of our experiment) and we considered the dependence from $\eta_R$. The opposite situation, where $\eta_R$ is kept fix is not reported. In this case $\Delta S_{\alpha}$ and $\Delta S'_{\alpha}$ behave similarly for all the variability range of $\eta_P$, and are always below $\Delta S_{\alpha}^{(unc)}$.

\begin{figure}[htbp]
	\centering
	\includegraphics[ trim= 1cm 0 0 0.5cm, clip=true,width=0.8\textwidth]{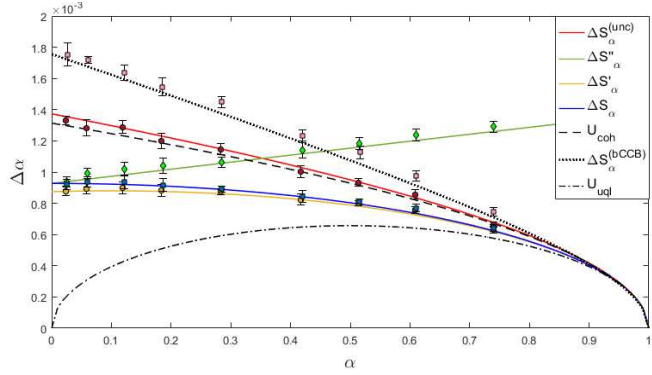}
	\caption{Uncertainty on $\alpha$ in function of the mean value of $\alpha$. Four different estimators are considered. Solid lines are the theoretical curves, dashed and dotted lines are the limits corresponding to the best quantum and classical cases, the markers are the experimental data. In this configuration, where measured efficiencies are $\eta_P=\eta_R=0.76$, the best sensitivity per photon ever achieved in loss estimation is demonstrated.}\label{graph}
\end{figure}

\begin{figure}[htbp]
	\centering
	\includegraphics[ trim=1cm 0 0 0.5cm, clip=true,width=0.8\textwidth]{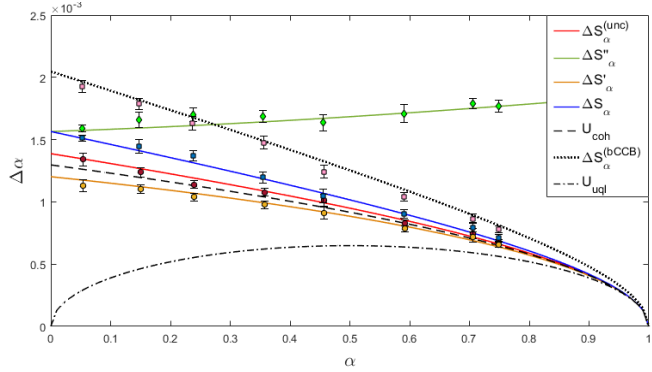}
	\caption{Uncertainty on $\alpha$ in function of the mean value of $\alpha$. Four different estimators are considered. Solid lines are the theoretical curves, dashed and dotted lines are the limits corresponding to the best quantum and classical cases, the markers are the experimental data. In this configuration the measured efficiencies are  $\eta_P=0.76$ and $\eta_R=0.43 < 0.5$. In this condition  $\Delta S_{\alpha} > \Delta S_{\alpha}^{(unc)}$  while $\Delta S$'$_{\alpha}$ remains below the classical benchmark.}\label{prova3lim}
\end{figure}

\begin{figure}[htbp]
	\centering
	\includegraphics[ trim= 1cm 0 0 0.5cm, clip=true,width=0.8\textwidth]{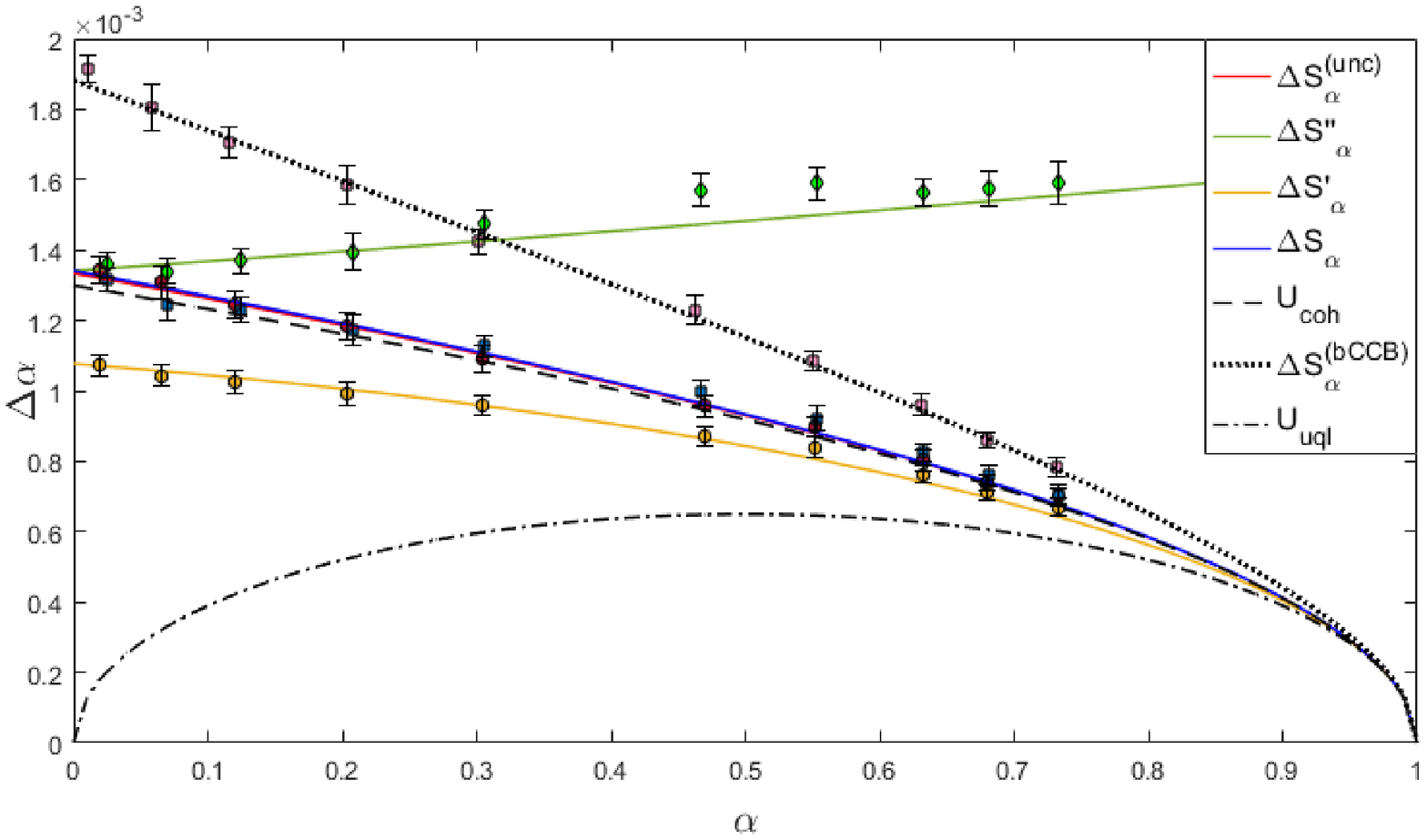}
	\caption{Uncertainty on $\alpha$ in function of the mean value of $\alpha$. Four different estimators are considered. Solid lines are the theoretical curves, dashed and dotted lines are the limits corresponding to the best quantum and classical cases, the markers are the experimental data. In this configuration the measured efficiencies are $\eta_P=0.76$ and $\eta_R=0.49$}\label{prova4lim}
\end{figure}

These different regimes at varying $\eta_R$ have been experimentally explored with our set-up and the results are shown in Fig. \ref{graph}-\ref{prova3lim}-\ref{prova4lim}.
In these figures, considering different estimators,  the dependence of the uncertainty on $\alpha$ in function of its mean value is reported. The three situations only differ from the value of $\eta_R$ considered. The solid lines are the theoretical curves in Eqs. (\ref{Ualpha_Uncorr}, \ref{Ualpha}, \ref{Ualpha_opt}-in the general case of $\eta_R \neq \eta_P$, \ref{aaa}) in which the experimental values of the quantities $\sigma_{\gamma}, F_P, F_R, \gamma$ have been introduced, while the markers represent the experimental data obtained. The black lines stand for significant limits, achievable in the ideal situation of unitary efficiencies on the two channels: the dotted-dashed line is the fundamental quantum limit $U_{uql}=[\alpha (1-\alpha) / \langle N_P \rangle]^{1/2}$, the dashed line is the classical limit in the direct case, $U_{coh}$, while the dotted line is the classical limit in the two-mode balanced case, $\Delta S_{\alpha}^{(bCCB)}$.
The good agreement between the theoretical curves and the experimental data witnesses the performances of our theoretical model in describing experimental imperfections.
Moreover, although not unitary efficiencies (in the best case, reported in Fig.\ref{graph} we measured $\eta_P=\eta_R= 0.76$) lead to a remarkable detachment from the UQL, we demonstrated the best quantum enhancement ever achieved in loss estimation experiments.
In particular, for $\alpha \sim 2\%$, we obtained $\Delta S'_{\alpha}/\Delta S_{\alpha}^{(unc)} =(1.51 \pm 0.13)$ and $\Delta S'_{\alpha}/\Delta S_{\alpha}^{(bBBC)} = (2.00 \pm 0.16)$.

The comparison with the two-mode classical strategy is of particular interest since the two-beam approach allows compensating unavoidable drifts and instability of source and detectors, leading to an \textit{unbiased} estimation of $\alpha$.
Using the two beams estimators $S_{\alpha}$ or $ S''_{\alpha}$ the only condition for having an unbiased estimator is that the ratio $\gamma = \frac{\langle N_P \rangle}{\langle N_R \rangle}$, evaluated in absence of the object, is stable. Experimentally, this is much less demanding than the stability of the probe beam (i.e. $\langle N_P \rangle$ constant over time) for the direct/single beam case. It is expected that the factors affecting the source and the detectors act in the same way on the probe and on the reference channels.

Another important point is to what extent the absorption measurement can be considered \textit{absolute}, i.e. not grounded on pre-calibration of the system against calibrated standard. Concerning $S_{\alpha}$ and $S''_{\alpha}$ they do not require any knowledge about the power of the source or the independent evaluation of the efficiencies of the detectors. The only parameter needing to be evaluated in a calibration phase is $\gamma$. On the other side $S'_{\alpha}$, in particular to calculate $k$ and $\delta E$, requires the knowledge of the two absolute values of both the efficiencies $\eta_R$ and $\eta_P$, virtually without uncertainty. Even if, in principle, they can be determined from the same set-up by using some extensions of the Klyshko's method \cite{model,brida2010OE,calibrazione} (i.e. as described in the previous section they can be extracted from the measured value of $\sigma_{\gamma}$) this could become cumbersome and it affects the final accuracy; in particular it requires firstly a long enough time to reduce the uncertainty to a negligible level and secondly a stability of the system from the characterization stage to the true measurement stage.

\section*{Conclusion}

We address the question of loss estimation and propose a simple measurement strategy, exploiting quantum correlations in twin beam which approach the ultimate quantum limit of the sensitivity in case of perfect detection efficiency. The experiment reports the best sensitivity per photon ever achieved in loss estimation without any kind of data post-selection, and confirms the theoretical model accounting for detection losses. In particular we double the sensitivity of a classical two-beam approach used in conventional measurement and we  overtake the best classical strategy, which uses shot noise limited beam, of more than 50\% (we consider the ratio of the standard deviations).

Our proposed measurement, represented by $S_{\alpha}$ in Eq.(\ref{S_alpha}), is compared both theoretically and experimentally, with other estimators in literature (see Eq.(\ref{opt}) and (\ref{S2})) in presence of experimental imperfections (e.g. not unitary detection efficiency). Despite in case of high detection losses the estimator $S'_{\alpha}$ in Eq.(\ref{opt}) has the smallest uncertainty, it turns out that where the quantum enhancement is significant, i.e. for sufficiently high efficiencies, $S_{\alpha}$ and $S'_{\alpha}$ approximately offer the same quantum enhancement. Moreover, we argue that  our procedure has several practical advantage, on one side being robust to experimental unavoidable drifts of the sources and detectors, leads to unbiased estimate. On the other side it does not need pre-characterization of the source noise and the detection efficiency calibration. In view of real applications these features are of the utmost importance and must be taken into account.

\section*{Acknowledgments}
The Authors thank I.P. Degiovanni, S. Pirandola and C. Lupo for elucidating discussion and N. Samantaray for his help in setting the preliminary phase of the experiment. 

\section*{Author contributions statement}

IRB and AM conceived the idea of the experiment, which were designed and discussed
with input from all authors. IRB and EL developed the theoretical model. EL and AM realized the experimental
setup and collected the data in INRIM quantum optics labs (coordinated
by MG). All authors discussed the results and contributed to the writing
of the paper.  All authors reviewed the manuscript.

\section*{Additional information}

\textbf{Competing financial interests} There is no competing financial interest.


%
%

\end{document}